\documentclass[%
 aip,
 amsmath,amssymb,
 reprint,%
]{revtex4-1}

\usepackage{graphicx}
\usepackage{dcolumn}
\usepackage{bm}

\usepackage[utf8]{inputenc}
\usepackage[T1]{fontenc}
\usepackage{mathptmx}
\usepackage{etoolbox}

\makeatletter
\def\@email#1#2{%
 \endgroup
 \patchcmd{\titleblock@produce}
  {\frontmatter@RRAPformat}
  {\frontmatter@RRAPformat{\produce@RRAP{*#1\href{mailto:#2}{#2}}}\frontmatter@RRAPformat}
  {}{}
}%
\makeatother
\begin{document}

\preprint{AIP/123-QED}

\title{CATALOGING THE NONLINEAR WAVES EXCITED BY MOVING A CHARGED BODY IN THE DUSTY PLASMA MEDIUM}
\author{Swathi S Krishna}
\affiliation{Department of Physics, Indian Institute of Science Education and Research, Dr. Homi Bhabha Road, Pune 411008, India}%

\author{S. K. Mishra}
 
\affiliation{Planetary Sciences Division, Physical Research Laboratory, Ahmedabad, 380009, India}%

\date{\today}
\author{S. Jaiswal}%
 \email{surabhijaiswal73@gmail.com}
\affiliation{Department of Physics, Indian Institute of Science Education and Research, Dr. Homi Bhabha Road, Pune 411008, India}%

\begin{abstract}
The nonlinear waves excited by the movement of a charged body in the dusty plasma medium are studied. A charged body moving through a dusty plasma medium can generate diverse nonlinear waves, such as precursors and pinned solitons. These wave excitations under weakly nonlinear and dispersive limits are described theoretically by the forced Korteweg-de Vries (fKdV) type equation. We have examined the role of the driver in shaping and evolving these wave excitations. In particular we studied the effect of primarily three source parameters, namely, amplitude, width, and flow speed, on the evolution of nonlinear structures.
The driver generates a perturbation in the stable system configuration, which couples with medium characteristics and eventually evolves into propagating excitations. Our finding shows that the excitation of nonlinear structure by a moving body in a plasma medium is not just dictated by the mach number but also the features of the source such as amplitude and width. As a novel finding apart from pinned and precursor solitons, we observe another nonlinear structure that lags behind the source term, maintaining its shape and speed as it propagates. These features are the first ever theoretical depiction of such lagging structures.

\end{abstract}

\maketitle

%

\section{\label{sec:introduction}Introduction}

Due to its potential uses and detectability, the occurrence of the linear and nonlinear wave structures like the wakes, precursor and pinned solitons which are excited by the motion of a charged object in a dusty plasma medium has been studied extensively in several investigations\cite{r1,r2,r3,r4,r5,r6,r7,r8,r9,r10,r11,r12}. These nonlinear perturbations can be potentially used to record small space debris objects in the Earth's ionosphere. These phenomena have been studied in several hydrodynamic \cite{r13,r14,r15,r16} studies. The reason for the formation of these structures is evident from those observations. When a moving disturbance travels through a medium of fluid, small-amplitude perturbations generate linear wakes—oscillatory waves formed by the superposition of linear modes. As the disturbance strength increases, nonlinear effects balance the medium’s dispersion, producing solitary waves (solitons) that maintain their shape while propagating independently. Thus, wakes arise from linear wave interference, while solitons emerge from nonlinear–dispersive balance.  These solitons are solutions to nonlinear partial differential equations and are characterized by their ability to maintain their shape and velocity as they propagate. The fascinating property of these structures is that the product of the amplitude and the square of the width remains constant while propagation.\\

The formation of nonlinear ion acoustic solitons has been predicted by Sen \textit{et al} \cite{r1}. They put forward the relevance and impact of these waveforms for the dynamical study of orbital space debris.
Precursor solitons travel ahead of the charged moving object at a higher speed than the moving object in the upstream direction, and the pinned solitons remain attached to the moving source term. These solitonic structures have been modelled using the forced Korteweg-de Vries (f-KdV) equation\cite{r1,r2}. Later, the first experimental demonstration of precursor solitons was given by Jaiswal \textit{et al}\cite{r3}. They observed the formation of precursor solitons upstream and the wakes downstream in a complex (dusty) plasma medium. Dusty plasma is composed of massive and highly charged micrometer or sub-micron particles in addition to the conventional plasma consisting of electrons, ions, and neutrals. 
 
Arora \textit{et al} \cite{r5} observed the formation of pinned solitons when they increase the Mach number to more than 1.4. They showed that these solitons move with the same velocity as the driver, and seem pinned to the source. 

El Nabulsi \textit{et al} \cite{r17} studied the various nonlinear solitonic structures which were modelled using the nonlinear evolution equation. Various functions, including the exponential, hyperbolic and trigonometric forms, were used to derive solutions to these nonlinear evolution equations. The related parameters determine the properties of the solitonic structures formed. Tiwari \textit{et al} \cite{r18} used molecular dynamics simulations to show that a charged supersonic projectile in a strongly coupled dusty plasma can create nonlinear fore-wake excitations, specifically precursor solitons and dispersive shock waves. The nonlinear plasma perturbations act as a stand-in for small orbital debris, offering a novel avenue for their detection and characterization\cite{r19,r20,r21,r22}.
\par The results of previous experimental studies \cite{r3,r4,r5} and theoretical studies \cite{r1,r2,r6,r7,r23} show substantial agreement on the formation of these nonlinear structures. These structures were also modeled using the forced Boussinesq equation\cite{r24,r25,r26}. However, in all the studies conducted before, the Mach number was considered as a threshold parameter for excitation of these nonlinear structure and an extensive establishment of the conditions for these excitations is still absent. We conducted a detailed study on the three major input parameters that define the source term, namely the amplitude, width, and velocity, to observe the formation of wakes and solitons. Initially, we solved the fKdV equation for the dust acoustic case using an analytical approach and subsequently performed the parametrization to catalog the formation of nonlinear structure. 

By our analysis, we were also able to identify a structure that lags behind the source term, essentially a solitonic structure. Unlike wakes, which move in the downstream direction, these lagging structure moves towards the source but lag behind.  Wake structure is  a dispersive wave packet which decays as time passes and distances from the source. The amplitude of a wake is not a constant but it's a function of position relative to the moving object and time. The lagging structure that is formed is not decaying over time, and it is also moving towards the source, whose amplitude increases to an extent and remains constant after a critical point. Our results are the first-ever theoretical depiction of these lagging structures.
\par This paper is organised as follows. Sec \textbf{II} contains the analytical solution of the forced Korteweg de-Vries equation. We then discuss how we solved it by using the pseudo-spectral method and subsequently substituted a wide range of values for the three parameters in Sec\textbf{III}.In Sec\textbf{IV} we discuss the observations of our study and arrive at the results. Sec\textbf{V} provides a brief summary and some concluding remarks. 
\section{Model fKdV  Evolution Equation}
The forced Korteweg-de Vries (fKdV) model equation for a dust acoustic medium is solved to gain a subjective understanding of the occurrence of the nonlinear wave structures in a dusty plasma. We have referred to the  theoretical model for fKdV equation for a charged object moving through a dusty plasma medium . Below we briefly describe forced KdV model for dusty plasma medium. The system is considered as a weakly collisional and nonmagnetized, consisting of electrons, ions, and negatively charged particles of dust. Here, both the lighter electrons and ions are presumed to follow the Boltzmann distributions with temperatures $T_e$ and $T_i$ correspondingly; further, the fluid equations describe the dust particles. The following are the one-dimensional fluid equations that govern the behavior of dust.
\begin{align}
\frac{\partial n_d}{\partial t} + \frac{\partial (n_d u_d)}{\partial x} &= 0, \\
\frac{\partial u_d}{\partial t} + u_d \frac{\partial u_d}{\partial x} &= \frac{\partial \phi}{\partial x}, \\
\frac{\partial^2 \phi}{\partial x^2} - n_d - a_e e^{\sigma \phi} + a_i e^{-\phi} &= S\left(x - v_d t\right).
\end{align}
These are the continuity equation,  momentum equation, and Poisson Equation, respectively.  Here, $n_d$, $u_d$ serve as the density and velocity of the dust fluid, and $\phi$ is the electrostatic potential. $a_e$ and $a_i$ are the constants which are defined as  \[a_e = \frac{n_{e0}}{Z_d n_{d0}}, \quad
a_i = \frac{n_{i0}}{Z_d n_{d0}}\]
where \( Z_d \) is the number of charges that are present on the dust particles' surface  and \[\sigma = \frac{T_i}{T_e}\]The charge density source term emanating from the charged item traveling at a speed $v_d$ from the fluid's frame is represented by the term \( S(x - v_d t) \). 
The given equations are normalized using:
\begin{align*}
x &\rightarrow \frac{x}{\lambda_D}, &
t &\rightarrow t \, \omega_{pd}, &
\phi &\rightarrow \frac{e \phi}{k_B T_i}, \\
n_d &\rightarrow \frac{n_d}{n_{d0}}, &
u_d &\rightarrow \frac{u_d}{C_D}.
\end{align*}
where \[\lambda_D = \left( \frac{k_B T_i}{4\pi n_{d0} Z_d e^2} \right)^{1/2}\]is the Debye length,\[\omega_{pd} = \left(\frac{4\pi n_{d0}Z_d^2 e^2}{m_d} \right)^{1/2}\]is the dust plasma frequency, and\[C_D = \left( \frac{Z_d k_B T_i}{m_d} \right)^{1/2}\]
is the dust acoustic speed. 
\par Using the standard reductive perturbation technique \cite{r1,r27} the dependent variables were expanded in a power series in terms of a small expansion parameter $\epsilon$, which is the measure of the smallness of the perturbed amplitude to the corresponding
equilibrium quantity.

By equating the same order terms of $\epsilon$ for the 1st and 2nd order, we obtain the fKdV equation as,
\begin{equation} 
\frac{\partial \phi^{(1)}}{\partial \tau} 
+ b \, \phi^{(1)} \frac{\partial \phi^{(1)}}{\partial \xi} 
+ B \frac{\partial^3 \phi^{(1)}}{\partial \xi^3} 
= B \frac{\partial S_2}{\partial \xi}.
\end{equation}

The coefficient  \begin{equation} 
B = \frac{v_{\rm ph}^3}{2}
\end{equation}
The  coefficient 
\begin{align}
b &= -\frac{v_{\rm ph}^3}{(\delta - 1)^2} 
\Biggl[ 
\delta^2 + (3 \delta + \sigma_i)\sigma_i 
+ \frac{1}{2} \delta (1 + \sigma_i^2) 
\Biggr]
\end{align}

The equation (7) is known as the fK-dV equation, where \[S_2(\xi - F t)\] is the moving charge source term, which is traveling at the speed of  
\begin{align*}v_d & \text{ (drift velocity)}, and\\F &= 1 - v_d.\end{align*}

Solitons provide a precise analytical solution for the KdV soliton\cite{r28}.  We carried out a numerical investigation of(4) for an arbitrary negative Gaussian source given by 
\[
S_2 = A e^{-(\xi + Ft/G)^2}
\] where A and G are constants determining the amplitude and width of the source and  $F$=1-$v_d$.

\section{Numerical Solution and Methodology}
To analyze the weakly nonlinear and dispersive regime,  a multiple-scale perturbation technique was employed, introducing stretched variables to separate fast linear waves from slow nonlinear evolution. This led to the derivation of a forced Korteweg de Vries (fKdV) equation, which captures the balance between nonlinearity, dispersion, and external forcing. The solutions of this equation reveal the formation of nonlinear solitons  and linear wakes.  Our analysis combines analytical derivations with numerical simulations to examine the generation and evolution of these nonlinear structures.

The  fKdV equation was studied and solved to obtain the analytical solution. We have employed the pseudo-spectral method to numerically solve this problem. The Fourier pseudo-spectral method \cite{r29} involves transforming the equation into Fourier space, where differentiation is replaced by multiplication, thereby simplifying computations. After solving in Fourier space, an inverse transform returns the solution to physical space.
We solved the forced Korteweg-de Vries (KdV) equation,
\begin{equation*}
a\,u_t + b\,u\,u_x + u_{xxx} = S_x,
\end{equation*}
on a periodic domain $x \in [0, L_x]$ using a Fourier pseudo-spectral method combined with an integrating factor scheme.
The solution $u(x,t)$ is discretized on $N$ uniformly spaced grid points, and spatial derivatives are computed exactly in Fourier space:
\begin{equation*}
\mathcal{F}[u_x] = i\,k\,\hat{u}_k, 
\quad
\mathcal{F}[u_{xxx}] = (i\,k)^3\,\hat{u}_k,
\end{equation*}
where $k = \frac{2\pi}{L_x}\left[0,1,\dots,\frac{N}{2}, -\frac{N}{2}+1, \dots, -1\right]$ are the wave numbers.

The nonlinear term 
\(-b\,u\,u_x\) 
can be rewritten as
\[
-b\,u\,u_x \;=\; -\frac{b}{2}\,\partial_x\!\left(u^{2}\right).
\]
This term is computed in physical space (by squaring \(u\)), and then transformed
back into Fourier space for further calculations.

The source term \(S(x,t)\) is prescribed explicitly, and its contribution
enters through its derivative \(S_x\), which is also evaluated spectrally.

In Fourier space, the governing equation takes the form
\[
\frac{d}{dt}\,\hat{u}^k
= \frac{1}{a}\Bigl( i k^{3}\,\hat{u}^k \;+\; \widehat{\text{Non}}^{\,k} \;+\; i k\,\hat{S}^{\,k} \Bigr),
\]
where \(\hat{u}^k\), \(\widehat{\text{Non}}^{\,k}\), and \(\hat{S}^{\,k}\) denote the Fourier transforms of the solution,
the nonlinear term, and the source term, respectively.

\vspace{0.5em}
To enhance temporal accuracy, a Richardson extrapolation strategy is employed.
At each time step, the solution is advanced both with:
\[
\text{(i)} \;\; \text{a single step of size } \Delta t,
\qquad\\
\text{(ii)} \;\; \text{two successive half-steps of size } \tfrac{\Delta t}{2}.
\]
The two results are then combined to yield an improved approximation.\\

Before investigating the importance of the input source parameters and medium nonlinearity in the formation of these nonlinear waves, the past study of Sen \textit{et al} \cite{r1}, Jaiswal \textit{et al}\cite{r3} and Arora \textit{et al} \cite{r5} is repeated to serve as a benchmark.\\
Sen \textit{et al} \cite{r1} have given the  exact analytical solutions for two different forms of $S_2$.\\
For \[
S_2 = A G \, sech^{2}\!\left[
\sqrt{\frac{A}{6}}
\left(\xi - \frac{2A - 3G}{6}\, t
\right)
\right]
\]
\[
\phi_1
= A\,sech^{2}\!\left(
\sqrt{\frac{A}{6}}
\left[
\xi - \frac{2A - 3G}{6}\, t
\right]
\right)
\]
For\[
S_2 = A G \, sech^{4}\!\left[
\sqrt{\frac{A-G}{6}}
\left(\xi - \frac{A - G}{3}\, t
\right)
\right]
\]
\[
\phi_2
= A\,sech^{2}\!\left(
\sqrt{\frac{A-G}{6}}
\left[
\xi - \frac{A - G}{3}\, t
\right]
\right)
\]
They have obtained the numerical solutions by giving a finite perturbation in $\phi$ as an initial condition and also given that these solutions can be obtained from a zero initial condition,where the driver generates the perturbations, which then develop to produce  soliton .We are primarily concerned about the effect of the moving source.The simulations start from zero initial conditions, and the perturbations are generated by the moving source term. Here, we used a Gaussian profile.  We then changed the three input  source parameters over a wide range of values. This helped us determine the relevance of each parameter, which in turn facilitates the formation of different types of nonlinear structures. Boundary conditions are chosen to minimize reflections or to mimic a large, open plasma system. The system evolves up to the desired final time, and snapshots of the solution are extracted at prescribed intervals for further analysis. Periodic boundary conditions are inherently satisfied due to the spectral representation.\\

\begin{table}[h]
\caption{Simulation parameters used for the analysis.}
\label{tab:parameters}
\centering
\begin{tabular}{ll}
\hline\hline
Parameter & Values \\ \hline
Nonlinearity coefficient ($b$) & 6.2,3.3,2 \\
Flow velocity($v_d$) &  1.5, 2, 3, 5 \\
Amplitude ($A$) & 0.5, 2, 6, 10 \\
Width ($G$) & 0.5, 1, 2, 4, 6,8 \\ 
\hline\hline
\end{tabular}                                                       
\end{table}

The study was carried out by varying three independent parameters over a wide range of values, as summarized in Table~\ref{tab:parameters}.

The methodology was as follows:
\begin{enumerate}
    \item For a fixed nonlinearity and a fixed velocity, and for a constant amplitude, the width $G$ was varied from its lowest to highest value.
    \item The same procedure was repeated for all amplitudes, keeping the width fixed.
    \item Steps (1) and (2) were repeated for all velocities.
    \item The whole analysis was carried out for Nonlinerity coefficients 3.3 and 2 in addition to 6.2.
\end{enumerate}
\begin{figure}[h!]
    \centering
    \includegraphics[width=1\linewidth]{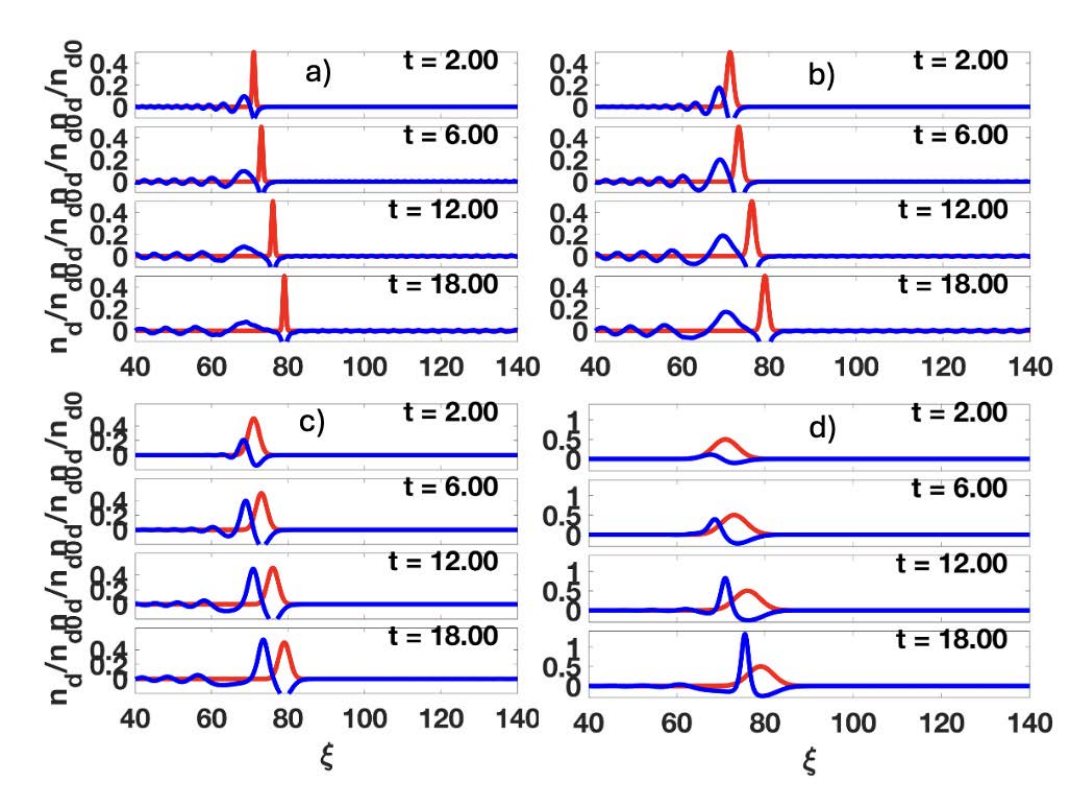} 
    \caption{Keeping amplitude constant and increasing width.For velocity 1.5,Nonlinearity 6.2,Amplitude 0.5 
  (a) wakes [$ G=0.5$], 
    (b) wakes [$ G=1$], 
    (c) lagging structure [$ G=2$], 
    (d) pinned [$ G=4$].}
    \label{fig:1}
\end{figure}

\begin{figure}[h!]
    \centering
    \includegraphics[width=1\linewidth]{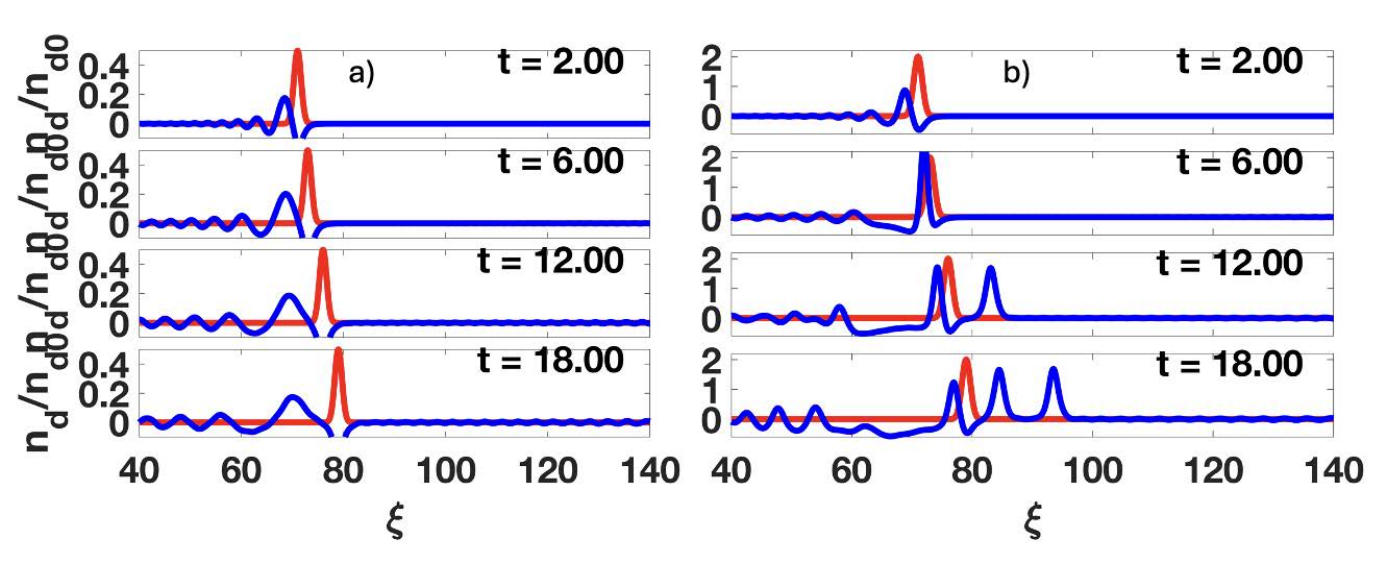} 
    \caption{Keeping width constant and increasing amplitude.For velocity 1.5,Nonlinearity 6.2,width 1 
  (a) wakes [$ A=0.5$], 
    (b) precursor [$ A=2$].}
    \label{fig:2}
\end{figure}

\begin{figure}[h!]
    \centering
    \includegraphics[width=1\linewidth]{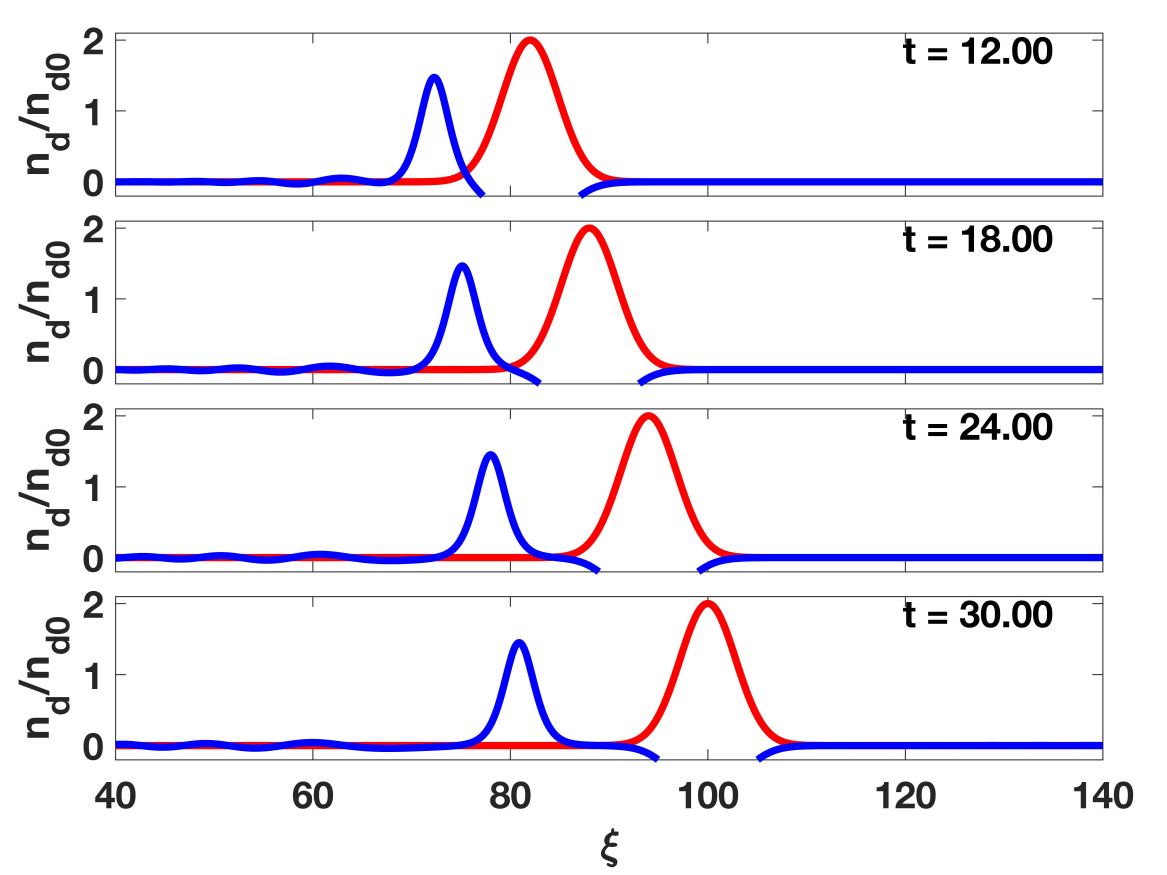} 
    \caption{Lagging structure formed for velocity 2;Amplitude 2;Width 4;Nonlinearity 2.}
    \label{fig:3}
\end{figure}
\begin{figure}[h!]
    \centering
    \includegraphics[width=1\linewidth]{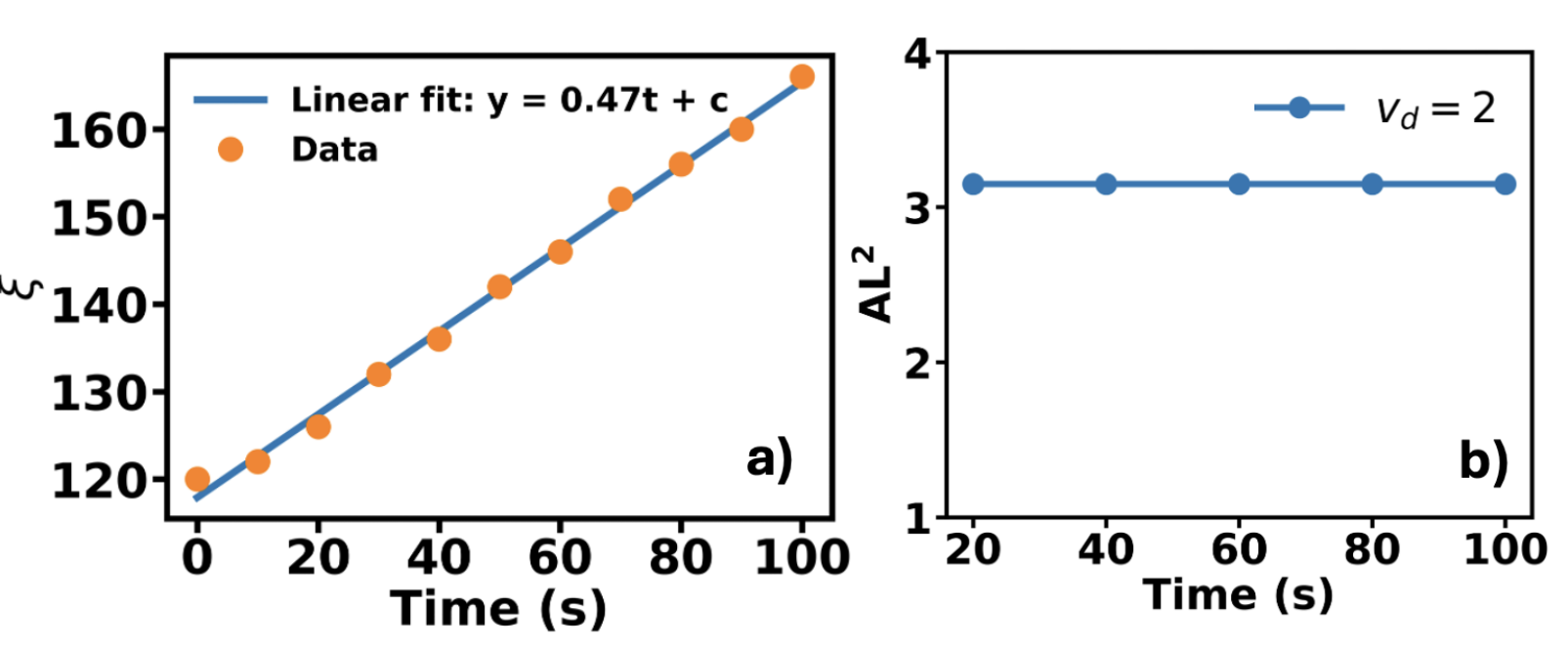} 
    \caption{
  (a) Velocity , 
    (b) AL$^2$ Value , for the lagging structure formed with ;A=2,G=4,vd=2,b=2.}
    \label{fig:4}
\end{figure}

\begin{figure}[h!]
    \centering
    \includegraphics[width=1\linewidth]{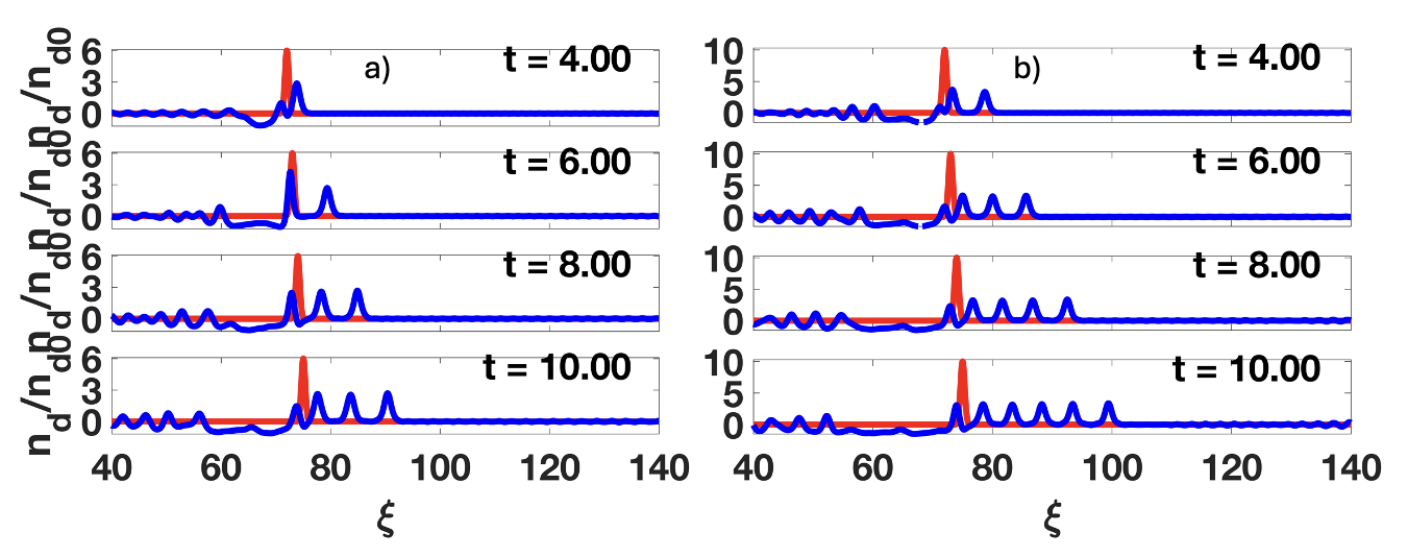} 
    \caption{Keeping width constant and increasing amplitude.For Velocity 1.5,Nonlinearity 6.2,width 0.5 
  (a) Precursor [$ A=6$], 
    (b) Precursor [$ A=10$].}
    \label{fig:5}
\end{figure}

\begin{figure}[h!]
    \centering
    \includegraphics[width=1\linewidth]{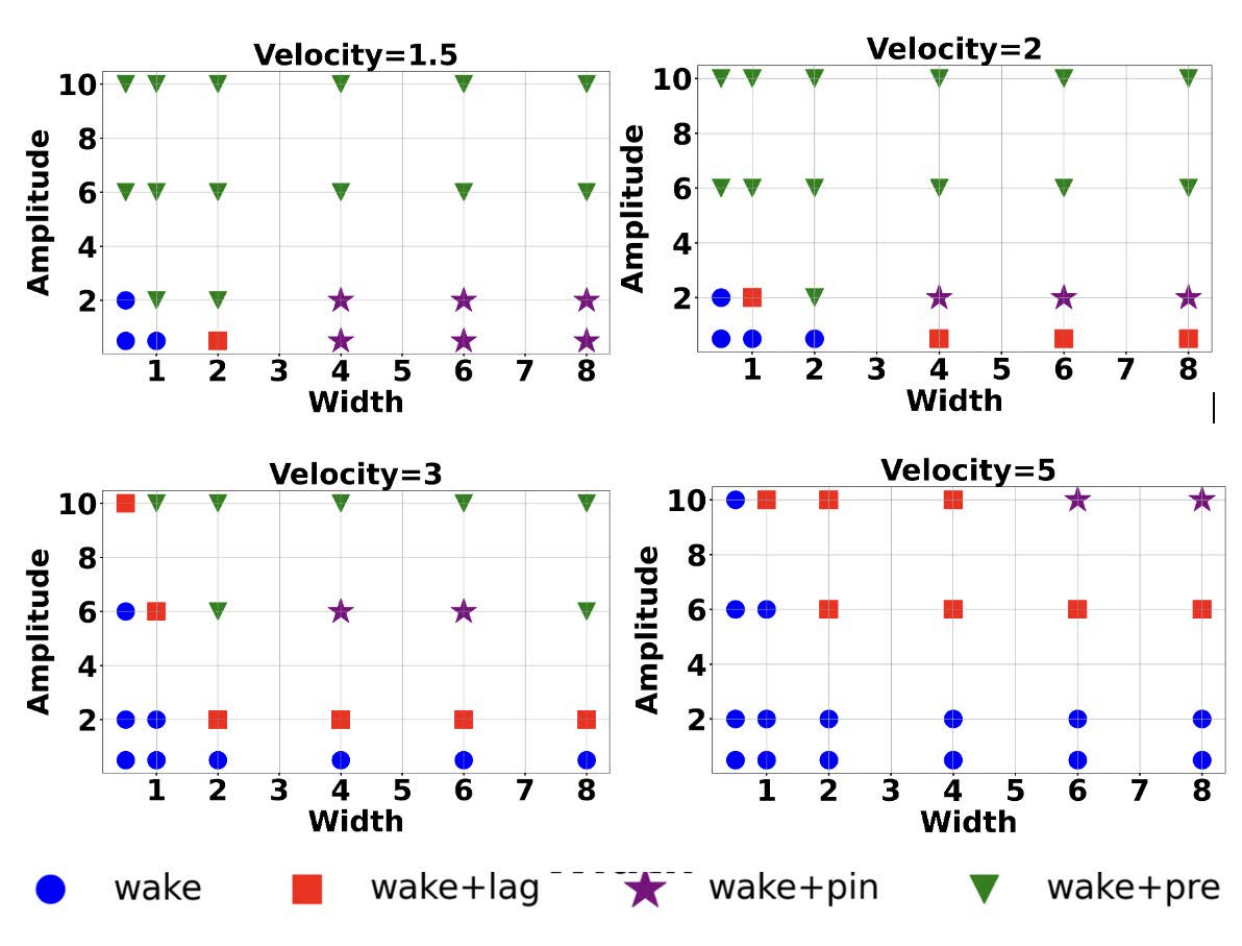} 
    \caption{Wave type distribution for Nonlinearity 6.2.}
    \label{fig:6}
\end{figure}

\section{Results and discussion}
By solving the fKdV equation, the evolution of NL structures is illustrated as a function of source parameters (viz., velocity, width, amplitude) and medium nonlinearity (\textit{b}) in Figs. 1-6. Coupling between the source and medium properties eventually leads the system to operate in linear and nonlinear regimes. Operating in the linear regime, the excitations primarily grow as wake field structures. In the nonlinear regime, the collective excitations grow as solitary waves in the upstream region, apart from weak dispersive excitations (wake fields) in the downstream region. 

Fig.1 depicts that the excitations evolve with varying width (G) from a linear wake (G = 0.5) to a pinned soliton (G = 4.0) that is mediated by a lagging structure (G = 2.0)  for vd = 1.5, b = 6.2, and A = 0.5. We observed that the pinned structure remains intact for increasing amplitude (A) as well, until a threshold value of A = 2. For higher amplitudes, the NL structures evolve into precursor solitons. The transition from linear wake oscillations to the precursor with increasing amplitude, while maintaining a constant width (G) and source velocity (vd = 1.5), is shown in Fig.2. A visual depiction of the evolving lagging structures is presented in Fig.3, which corresponds to vd = 2, b = 2, G = 4, and A = 2. It is noted that the lagging structure maintains its shape and size (i.e., A*L*L) as it propagates through the medium following the source, though with a lower speed (Fig~\ref{fig:4}, right panel) - this very much resembles the characteristics of solitonic structures. Moreover, the lagging structure is observed to propagate at a constant speed, as depicted in Fig.~\ref {fig:4} (left panel). Fig~\ref{fig:5} depicts the generation of precursor solitons with increasing amplitude - both the number of emitted precursor solitons and their velocity are observed to increase. This trend is consistent with the earlier findings reported by Arora \textit{et al} \cite{r4}.

With an understanding of the excitation and evolution of linear and nonlinear structures, and their dependence on key source and medium parameters, it is of interest to define a parameter space that describes the generation of specific phenomena/structures. To illustrate this, the simulation is performed on a trial-and-error basis to constrain the parameters (A, G, b, vd), identifying the excitation thresholds for wake, lagging, pinned, and precursors. As an illustration, an amplitude-width (A-G) space is defined for four different values of vd, keeping the nonlinear term (b) constant, and is shown in Fig.6. 

The precursors dominate at higher amplitudes and smaller drift velocities, resulting in stronger coupling with the medium and more pronounced nonlinear excitations. For weaker amplitudes and higher drifts, the wake features dominate, whereas for moderate amplitudes, lagging structures are excited. The pinned solitons exhibit between lagging and precursor excitations within the parameter space. The transition between these structures can clearly be seen left bottom panel (vd = 3). For instance, for a given amplitude A = 6, the figure illustrates the NL structure transitions from wake to precursor, mediated by lagging and pinned structures, with increasing width; a similar feature is observed with increasing amplitude for a given width (G = 4). With increasing drift, a relatively weaker coupling with the medium yields a wake-dominated region at lower amplitudes, while higher amplitudes are populated by pinned and lagging structures; the precursor threshold may occur at higher amplitudes. In the presentation, the illustration is made only for b = 6.2; however, calculations for other values of the nonlinearity parameter are also performed, and additional comparative figures with extended analyses are included in the supplementary material.

\section{\label{sec:conclusion}Conclusion}
To conclude, the possible nonlinear excitations due to a moving charged body in a dusty plasma medium are investigated and cataloged as a function of source and medium characteristic properties. The fKdV treatment is used to study the phenomenon and is detailed explored through both analytical and numerical approaches. We note that the driver,i.e., the moving source, allows perturbations that propagate and induce linear wakes and nonlinear soliton structures. As a novel finding apart from pinned and precursor solitons, we discover another nonlinear structure that lags behind but follows the source. Unlike wakes, which move downstream and decay over time, these lagging structures move towards the source. Initially, their amplitude increases over time to a certain limit, then remains constant, much like solitons. These lagging structures mediate the transition of excited perturbations from the linear wake to pinned and precursor solitons - a parametric space describing such a transition is illustrated as a function of source and medium properties. It gives the impression that such a transition is a sequential feature of the evolution of perturbative excitations that triggers, depending on a suitable regime of parametric operation. Such nonlinear excitations are not only of fundamental importance but also have significant practical implications. These nonlinear excitations may provide a suitable coherent electrical signature that, in principle, can be detected by ground-based or in situ measurements, and may provide a useful means of tracking orbital debris objects. 

\begin{acknowledgments}

\end{acknowledgments}

 \section*{Data Availability Statement}
The data that support the findings of this study are available from the corresponding author upon reasonable request.

\nocite{*}
\bibliography{aipsamp}

\end{document}